\documentclass[aps,prl,twocolumn,showpacs,amsmath,amssymb,superscriptaddress]{revtex4}
\usepackage{dcolumn}
\usepackage{bm}
\usepackage{graphicx}
\usepackage{epstopdf}

\def\bvpf{{\mathbf{p}_{\rm f}}}
\begin{document}

\renewcommand{\section}[1]{{\par\it #1.---}\ignorespaces}

\title{Model evidence of a superconducting state with a full energy gap in small cuprate islands}
\author{Annica M. Black-Schaffer}
\affiliation{Department of Physics and Astronomy, Uppsala University, Box 516, S-751 20 Uppsala, Sweden}
\author{Dmitri S. Golubev}
\affiliation{Institute of Nanotechnology, Karlsruhe Institute of Technology (KIT), Herman-von-Helmholtz-Platz 1, 76344, Eggenstein-Leopoldshafen, Germany} 
\author{Thilo Bauch}
\affiliation{Department of Microtechnology and Nanoscience, Chalmers University of Technology, S-412 96 G\"oteborg, Sweden}
\author{Floriana Lombardi}
\affiliation{Department of Microtechnology and Nanoscience, Chalmers University of Technology, S-412 96 G\"oteborg, Sweden}
\author{Mikael Fogelstr\"om}
\affiliation{Department of Microtechnology and Nanoscience, Chalmers University of Technology, S-412 96 G\"oteborg, Sweden}
\date{\today}

\begin{abstract}
We investigate subdominant order parameters stabilizing at low temperatures in nano-scale high-T$_c$ cuprate islands, motivated by the recent observation of a fully gapped state in nanosized YBa$_2$Cu$_3$O$_{7-\delta}$ [D. Gustafsson {\it et al}, Nature Nanotech.~{\bf 8}, 25 (2013)]. Using complementary quasi-classical and tight-binding Bogoliubov-de Gennes methods, we show on distinctly different properties dependent on the symmetry being $d_{x^2-y^2}+i s$ or $d_{x^2-y^2}+i d_{xy}$. We find that a surface-induced $d_{x^2-y^2}+i s$ phase creates a global spectroscopic gap which increases with applied magnetic field, consistent with experimental observation.
\end{abstract}
\maketitle

%
It is well established that high-temperature cuprate superconductors have a dominantly $d_{x^2-y^2}$-wave order parameter symmetry 
\cite{vanHarlingen1995, Tsuei2000}, but the existence of a subdominant symmetry has also long been considered. 
Of special interest are order parameters such as $d_{x^2-y^2}+id_{xy}$ ($d_1+id_2$) 
and $d_{1}+is$, which fully gap the Fermi surface and break time-reversal (${\cal{T}}$) symmetry \cite{Fogelstrom1997, Volovik1997, Laughlin1998, Balatsky1998}. Experimental data have been contradictory, invoking large imaginary subdominant orders to explain tunneling experiments in 
YBa$_2$Cu$_3$O$_{7-\delta}$ (YBCO) \cite{Covington1997, Elhalel2007} and La$_{2-x}$Sr$_{x}$CuO$_4$ \cite{Gonnelli2001} or thermal conductivity in Bi$_2$Sr$_2$CaCu$_2$O$_8$ \cite{Krishana1997} while, on the other hand, only very small imaginary components seem compatible with the absence 
of any measured spontaneous magnetization \cite{Tsuei2000, Carmi2000,Neils2002, Kirtley2006, Saadaoui2011}. 

The possibility to find a ${\cal{T}}$-symmetry breaking state is enhanced if the dominant $d_1$-wave order parameter is locally reduced by e.g.~surface scattering. A nano-scale island of a curate superconductor is, by virtue of its large surface-to-area ratio, thus an ideal candidate to search for this elusive state. Very recently, an even/odd parity effect was reported in YBCO single-electron 
transistors (SETs), signalling a fully gapped low-temperature superconducting phase in nano-scale YBCO \cite{Gustafsson2012}. 
For a pure $d_1$-wave superconductor there are always low-energy quasiparticle states available at the nodal points, into which the added charge in a SET can relax.  
Unless the nodal quasiparticles states are lifted by a spectroscopic energy gap $E_g$, a 
parity effect should not be present in YBCO-SETs. The experimental data show $E_g\approx 20-40$~$\mu$eV, making $E_g$ three orders of magnitude smaller than the gap $\Delta_{d_1} \approx 20$~meV in YBCO \cite{Gustafsson2012}. 

The YBCO SETs studied in Ref.~\cite{Gustafsson2012} have sizes of order $0.2\times0.2\times 0.1~\mu{\rm m}^3$, making the energy-level spacing $\delta_s$ of the nodal quasiparticles a candidate for the observed gap $E_g$. While $\delta_s \sim 10$~$\mu$eV for an infinite mean-free path, surface disorder gives a reduced $\delta_s\sim 0.01$~$\mu$eV $\ll E_g$ \cite{Dimasnote}. 
However, energy level spacing due to finite size is incompatible with the finite onset voltages of the SET current measured over a full gate charge period~\cite{kouwenhoven1998}. 
The measured $E_g$  is thus of superconducting origin and gap the whole Fermi 
surface. Given the parent $d_1$-wave order parameter of cuprates, finding $E_g$ demonstrates the presence of a complex order parameter, which gaps the nodal $d_1$-wave spectrum.

The aim of this Letter is to establish which fully gapped superconducting state nucleates at low temperatures in nano-scale cuprate islands. Specifically, we focus on three particularly illuminating experimental results in Ref.~\cite{Gustafsson2012}: a) The spectral gap $E_g \approx 20-40$~$\mu$eV, b) the energy level spacing $\delta_s \sim 1$~$\mu$eV above $E_g$, and c) $E_g$ increases with applied magnetic field in the range 0-3~T. 
We use complementary quasi-classical and tight-binding Bogoliubov-de Gennes methods to show that the $d_1+is$ and $d_1+id_2$ states have distinctly different properties in small cuprate islands. 
While a subdominant $s$-wave component appears at the surface due to disorder suppression of the $d_1$-wave, a $d_2$ component can only nucleate in the interior of the island. Furthermore, the $d_1+is$ state has a finite gap set by the decay of the $s$-wave order into the center of the island, whereas the low-energy spectrum of the $d_1+id_2$ state is determined by finite size quantization of its chiral surface states. Thus, the $d_1+is$ state has a low-energy spectrum similar to that of a conventional $s$-wave superconductor, qualitatively satisfying points a) and b), while the $d_1+id_2$ state has equally spaced low-energy levels. We also find the magnetic field dependence to only be consistent with the $d_1+is$ state.
%
\begin{figure*}[t]
\includegraphics[scale = 0.9]{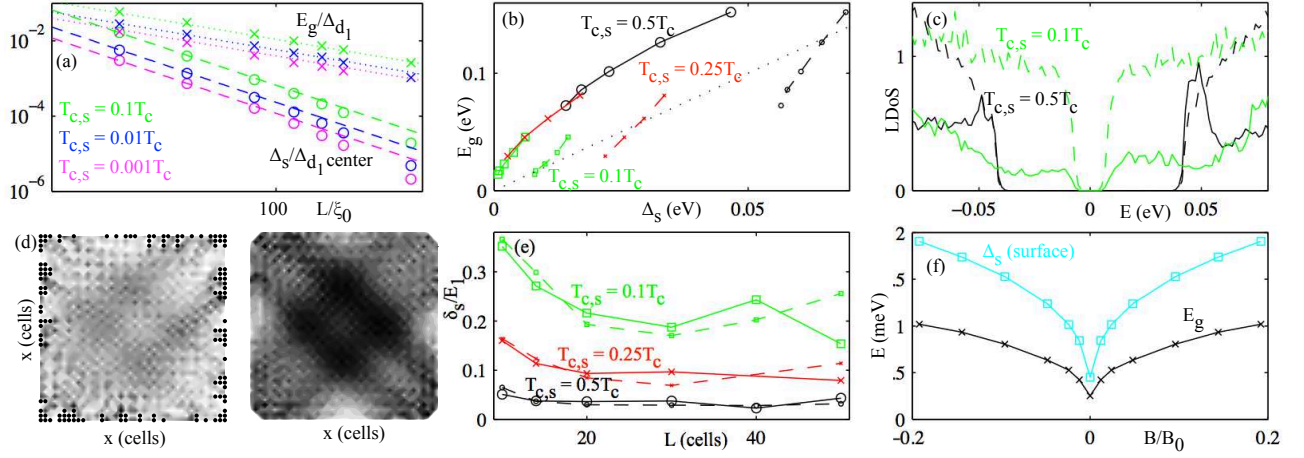}
\caption{\label{fig:DiS} (Color online) Subdominant $s$-wave order. (a): QC $s$-wave component $\Delta_s$ in the center of the island ($\circ$) and nodal energy gap $E_{g}$ ($\times$) of a pair-breaking [110] surface with disorder as function of slab length $L/\xi_0$ for $T_{c,s}/T_c = 0.001$ (magenta), 0.01 (blue), 0.1 (green) (increasing values) at temperature $T = 0.01T_c$. Dashed lines are $\Delta_s({\rm center})/\Delta_{d_1} = 4\Delta_s({\rm surface})/(L/\xi_0)$, dotted lines are $E_g/\Delta_{d_1} = 48\Delta_s({\rm surface})/[(1-\log(T_{c,s}/T_c))(L/\xi_0)^2]$.
(b): BdG disorder averaged energy gap $E_g$ as function of $\Delta_s$ in the center of the grain (solid) and at the surface (dashed) for $T_{c,s} = 0.5 T_c$ (black, $\circ$), $0.25T_c$ (red, $\times$), and $0.1T_c$ (green, $\square$). Dotted line is $E_g = 2\Delta$. (c): BdG disorder averaged LDoS in the center of the grain (solid) and at the surface (dashed). (d): BdG eigenstate spatial density for one disorder configuration with $L = 40$ (black dots marking removed sites) averaged over the 4 lowest energy states (left) and disorder averaged (40 configurations) for the lowest energy state (right). White = zero, black = 0.01 (left) or 0.005 (right) states per unit cell. (e): BdG disorder averaged level spacing $\delta_s = E_{n+1}-E_{n}$ ratio to first energy value $E_1$ for $n = 1$ (solid) and $n = 2$ (dashed) as function of grain size $L$. (f): Evolution of the QC energy gap $E_g$ (black, $\times$) and $\Delta_s$ on the surface (cyan, $\square$) with magnetic field, $B_0=\Phi_0/\pi\xi_0 \lambda_0\approx2.2$~T (derived assuming $\xi_0 = 2$~nm, $\lambda_0 = 150$~nm for YBCO) for  a $L=40~\xi_0$ slab with surface disorder modeled by a thin layer ($\sim0.2 \xi_0$) with a graded impurity concentration \cite{Laube2004} and $T_{c,s} = 0.065T_c$.
}
\end{figure*}
%

%
%
The necessary attraction in a subdominant pairing channel for a gapped state is an inherent feature of boson-mediated pairing, such as a spin-fluctuation model relevant for high-T$_c$ cuprates \cite{Radtke1992, Buchholtz1995, Eschrig2001, Fogelstrom2004, Fogelstrom2011}. 
The relative strength of the attraction in different paring-symmetry channels is sensitively dependent on the shape of the pairing susceptibility and the band structure in vicinity of the Fermi level. In small cuprate islands, with spatial dimensions down to 50-100 times the superconducting coherence length $\xi_0$, finite size effects might become important such that the pairing interactions deviate from the bulk values.
Here we will not further elaborate on this but instead focus on the consequences of an assumed subdominant pairing by following Refs.~\cite{Fogelstrom1997, Rainer1998} to generate generic phase diagrams for the $d_1+is$ and $d_1+id_2$ states. We quantify the strength of the subdominant $d_2$- or $s$-wave pairing simply by their bare bulk transition temperature $T_{c,d_2/s}$, measured relative to $T_c$, the transition temperature into the $d_1$ state.

To treat the appearance of subdominant orders in a cuprate island we employ two complementary methods. First we calculate the nucleation of subdominant order parameters in [100] and [110] surface slabs within the weak-coupling quasiclassical (QC) approximation for superconductivity (see e.g.~Ref.~\cite{ Eschrig2009}). 
The advantage of this approximation is that we can self-consistently compute the order parameter mean-fields in restricted geometries of realistic size, including effects of both single-impurity scattering, impurity self-energies, and surface scattering. A self-consistent evaluation of both thermodynamical properties, such as order parameter fields and magnetization, and the spectral properties is done as function of subdominant pairing strength and applied magnetic field. Since the QC approximation is a leading-order theory in quantities such as $1/(k_F\xi_0) \approx 0.1- 0.2$ for high-T$_c$ cuprates, it cannot, however, resolve effects of a finite level spacing $\delta_s$. 
To complement these results, we therefore also study two-dimensional (2D) square lattice islands within the Bogoliubov-de Gennes (BdG) framework \cite{deGennesbook}:
\begin{align}
\label{eq:HBdG}
& H_{\rm BdG}  =   \sum_{i,j,\sigma} t_{ij} c^\dagger_i c_j + \sum_{\langle i,j\rangle} \Delta_{d_1}(i,j)[c^\dagger_{i\uparrow} c^\dagger_{j\downarrow} - c^\dagger_{i\downarrow}c^\dagger_{j\uparrow}]  \\ \nonumber
& + \sum_{i}\! \Delta_s(i) c^\dagger_{i\uparrow}c^\dagger_{i\downarrow} + \sum_{\langle \langle i,j\rangle \rangle} \!\! \Delta_{d_2}(i,j)[c^\dagger_{i\uparrow} c^\dagger_{j\downarrow} - c^\dagger_{i\downarrow}c^\dagger_{j\uparrow}] + {\rm H.c.}.
\end{align}
Here we use a band structure parameterization $t_{ij}$ relevant for a 2D model of high-$T_c$ cuprates \cite{Hoogenboom2003}, including up to next-next nearest-neighbor hopping. 
The dominant $d_1$-pairing is implemented on nearest-neighbor bonds \cite{Tanuma1998} with the (mean-field) order parameter calculated self-consistently using $\Delta_{d_1} = -V_{d_1}\langle c_{i\downarrow} c_{j\uparrow} - c_{i\uparrow}c_{j\downarrow}\rangle$, with $\Delta_{d_1}(x)$ and  $\Delta_{d_1}(y)$ treated independently. Due to computational demands limiting our grain sizes, we set $V_{d_1} = 0. 455$~eV giving coherence peaks at energies 5 times larger than in YBCO \cite{Aprile1995}. Thus, while the BdG approach accurately captures the short $\xi_0$ in cuprates, we are instead limited by studying smaller grains and stronger superconductivity than found experimentally.
Subdominant $s$-wave pairing is introduced through a negative on-site potential $V_s$, with self-consistency equation $\Delta_s = -V_s \langle c_{i\downarrow} c_{i\uparrow} \rangle$. For subdominant $d_2$ pairing we use a next-nearest neighbor pair potential $V_{d_2}$ analogous to $V_{d_1}$ but with imposed $d_2$ symmetry.
We study islands with an overall [100] square shape with sides as large as $L = 50$ unit cells, thus $L/\xi_0$ is comparable to recent experiments \cite{Gustafsson2012}. We introduce surface disorder by randomly removing up to 25\% of the surface atoms in the three outermost surface layers, see Fig.~\ref{fig:DiS}(d). While no disorder average errorbars are displayed in the data in Figs.~1-2, they are negligible for all but the smallest islands. 
We find no significant difference in energy levels and subdominant orders between these disorder configurations and those of a diamond shaped [110] island with moderate surface disorder, thus establishing the generic nature of our islands.

\section{Subdominant $s$-wave order}
Surface scattering severely suppress $d$-wave order at generic surfaces. 
Using both the QC and BdG methods we find that, given any finite pairing interaction in the $s$-wave channel, an $s$-wave order parameter with relative phase $\pi/2$ nucleates in the $d_1$-voided disordered surface region, producing a ${\cal{T}}$-symmetry breaking $d+is$ state. $\Delta_s$ is constant in the surface region, only determined by the $s$-wave pairing strength.
The surface $s$-wave state leaks into the center of the island, with $\Delta_s {\rm (center)} \propto \Delta_s({\rm surface})/L^{2}$ for the experimentally relevant grain sizes.
We have analytically verified this power law decay by expanding the free energy of the bulk superconductor in
$\Delta_s$ to second order while keeping the dominating $d$-wave component $\Delta_{d_1}$ exactly. The correction to the  free energy takes the form
$\delta F_s = \int \frac{d^2q}{(2\pi)^2}\chi_s(q_x,q_y)\Delta_s^*(-\bm{q})\Delta_s(\bm{q})$, where the kernel
$\chi(q_x,q_y)\propto 1+\xi_s|q_x-q_y|+\xi_s|q_x+q_y|$ reflects the strong anisotropy of the quasiparticle dynamics,
and $\xi_s\sim V_s/\Delta_{d_1}$ is the $s-$wave coherence length. Accordingly, we find that, in agreement with the numerics,
far away from a long boundary located at $x=0$, the $s$-wave component 
should decay as $\Delta_s\sim \int \frac{dq_x}{2\pi} e^{iq_x L} \chi^{-1}(q_x,0)\sim \xi_s^2/L^2 $ \cite{Dimasnote}. 

Both the QC and BdG results have a finite gap $E_g$ in the energy spectrum, due to the finite $s$-wave order gapping the 
$d_1$ nodes. 
The QC spectral gap $E_g$ decays as $\Delta_s({\rm surface})/L$, as seen in Fig.~\ref{fig:DiS}(a). This scaling is a consequence of the Fermi-surface position ($\bvpf$) dependence of the effective coherence length, being very long in the nodal direction of a $d$-wave superconductor.
In Fig.~1(b) we see that the BdG $E_g$ for small islands depends monotonically on $\Delta_s$(center), largely independent on both island size and $s$-wave pair potential. 
However, the value of $E_g$ is much larger than the local result $E_g = 2\Delta_s$(center), because of the 
large $\Delta_s$(surface), and for larger islands the surface order parameter ultimately determines $E_g$. 
%
%
In Fig.~1(c) we plot the BdG local density of states (LDoS) in both the center and surface regions which 
shows that the gap is a global property of the island, only the LDoS above the gap is position dependent. The non-localized spatial density of the lowest energy eigenstates in Fig.~1(d) further cements the fact that the gap depends globally on the subdominant order parameter. 
Further focusing on the lowest energy levels and their spacings, we plot in Fig.~1(e) the ratio of $\delta_s$ to the lowest energy level $E_1 = E_g/2$. 
This ratio decreases with increasing $T_{c,s}$ for two reasons: first, $E_g$ increases and second, $\delta_s$ decreases due to a more pronounced 
coherence peak [see Fig.~1(c)] above the $s$-wave nodal gap. The low-energy spectrum of a cuprate nano-scale island with a subdominant $s$-wave state 
nucleated only on the surface is thus essentially that of a conventional $s$-wave superconductor. 
We find that in order to achieve $\delta_s/E_1 \sim 0.05$, 
as found experimentally \cite{Gustafsson2012}, $T_{c,s}\sim 0.5 T_c$ for $\Delta_{d_1}$ both 2.5 and 5 times larger than the experimental value. This 
indicates that the $s$-wave pairing might be strong close to the surface of nano-scale cuprate islands, although the energy gap is still very small, since 
$T_{c,s}\sim 0.5 T_c$ only supports a surface $s$-wave state.
Finally, in Fig.~1(f) we plot the QC magnetic field dependence for a surface disordered slab. $\Delta_s$(surface) grows with 
magnetic field and we see that $E_g$ 
closely tracks this behavior, consistent with experimental results \cite{Gustafsson2012}. 
\begin{figure*}[ht]
\includegraphics[scale = 0.9]{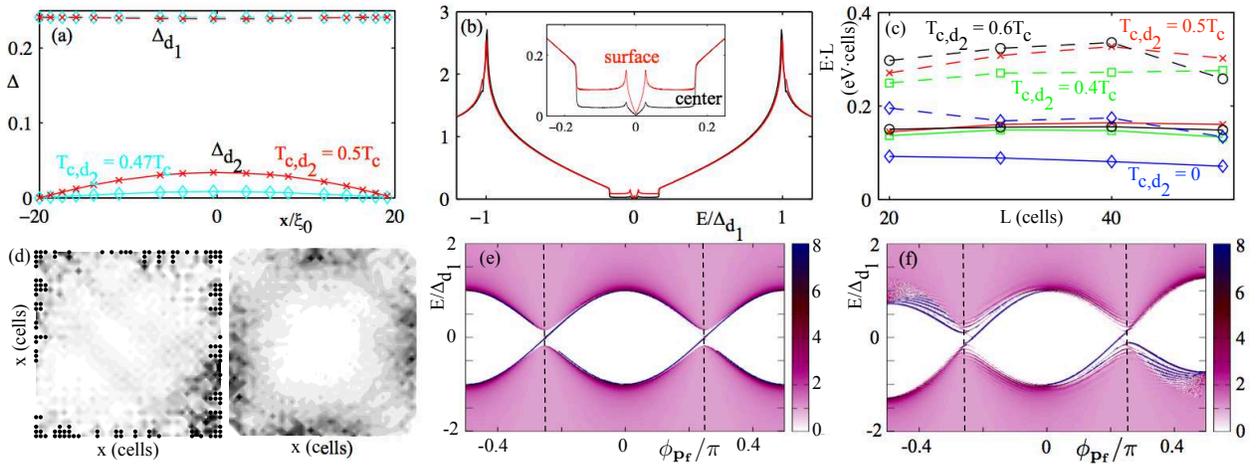}
\caption{\label{fig:DiD} (Color online) Subdominant $d_2$-wave order. (a): QC $d_1$-wave (dashed), and $d_2$-wave (solid) components across a $L = 40\xi_0$ [100] slab for $T_{c,d2} = 0.5T_c$ (red, $\times$) and  $0.47T_c$ (cyan, $\diamond$). (b): QC LDoS in units of normal state LDoS in the center (black) and at the surface (red). Inset shows a zoom-in at low energies. (c): BdG energy eigenstates times grain size $L$ as function of $L$ for $E_1$ (solid) and $E_2$ (dashed) for $T_{c,d2} = 0.6T_c$ (black, $\circ$), $0.5 T_c$ (red, $\times$), $0.4T_c$ (green, $\square$), and $0T_c$ (blue, $\diamond$).  (d): Same as Fig.~1(e) but for $d_2$-wave order. (e-f): QC band structure at the surface for $B = 0$ (e) and $B = 2B_0$ (f) with color scale showing the angle resolved LDoS, i.e $\phi_\bvpf$ measures the position on the Fermi circle measured from the $k_x$-axis. Dashed lines mark the $d_1$ node.
 }
\end{figure*}
%

%
\section{Subdominant $d_{xy}$-wave order}
While $s$-wave order survives significant disorder, that is not the case for $d_2$ subdominant pairing. We find using both the QC and BdG methods that a $d_2$ component nucleates only away from the surface, see Fig.~2(a), and then with a $\pi/2$ relative phase shift. We find that $T_{c,d_2} \gtrsim 0.4 T_c$ is needed for a $d_1+id_2$ state \cite{Rainer1998}, which requires an enhancement of the pairing attraction in the $d_2$-wave channel due to finite size effects. We will here not further discuss the probability of small dimensions increasing $T_{c,d_2}$, but instead assume this can be the case and focus on the consequences.  
Any $d_1+id_2$ state, even with a very small $d_2$ component, hosts two chiral edge states \cite{Volovik1997, Laughlin1998, Senthil1999}, independent on the surface morphology. Fig.~2(b) shows how the surface states produce a constant LDoS in the surface region, due to their 1D Dirac spectrum, and how this density leaks into the middle of the island in the QC results. The finite density of surface states in the center produces a very small hybridization gap as evident in the DoS very close to zero energy. We estimate the level spacing in these chiral modes to be $\delta_s \approx \hbar v_g \delta k$ which is on the order of $10$~$\mu$eV for an island circumference of $500$~nm. Here the gap-velocity $v_g=2\Delta_{d_1}/\hbar k_{\rm{F}}$ measures how $\Delta_{d_1}$ opens in the node. In Fig.~2(c) we complement the QC result by plotting the lowest energy levels in the BdG results. Notably, they scale as $L^{-1}\sim \delta k$, are independent of $T_{c,d_2}$, and are equally spaced, i.e. $\delta_s \approx E_1$. These results all confirm that the lowest energy states in a $d_1+id_2$ superconducting island are those of the chiral edge states. This is also evident from the spatial density of the lowest energy states shown in Fig.~2(d). 
The only possibility of avoiding measuring the chiral edge states would be in small islands with $\delta_s>E_g$, where $E_g$ is the nodal gap in the center of the island. However, we see in Fig.~2(c) that even with $T_{c,d_2} = 0$, the spectrum is still equally spaced, now from a finite size quantization in the $d_1$ nodes.
Both low-energy level spacings and eigenstate densities are thus distinctly different for a $d_2$ subdominant order compared to an $s$-wave order.
In Fig.~2(e, f) we examine the magnetic field dependence of the QC band structure. The main effect of the magnetic field is to shift the zero energy momentum of the edge modes. Thus, $\delta_s$ will not change with field. We also note that the QC magnitude of the subdominant $d_2$ component does not noticeably grow with magnetic field for fixed pairing strength.
Thus, neither energy level spacings nor magnetic field dependence of the $d_1+id_2$ state is seemingly consistent with current experimental data \cite{Gustafsson2012}. 
We finally note that self-consistent BdG results for both finite $s$-and $d_2$-wave pairing strengths, in general results in only one emergent subdominant order, despite their spatially separated nucleation regions, thus producing spectra similar to those already discussed.

\section{Spontaneous currents}
Both the $d_1+is$ and $d_1+id_2$ state generate spontaneous currents due to ${\cal{T}}$-symmetry breaking. The current experimental status regarding a superconducting state breaking ${\cal{T}}$-symmetry taken from scanning-SQUID experiments \cite{Kirtley2006} and $\beta$-detected nuclear magnetic resonance \cite{Saadaoui2011} put a strict upper limit on a subdominant order parameter $\Delta_{\rm{sub}}\lesssim 0.02\Delta_{d_1}$ \cite{Kirtley2006}. 
Also the spontaneous magnetization is limited by an upper bound of $0.2$~G \cite{Saadaoui2011}. 
Within the BdG framework we can calculate the quasiparticle currents by combining the charge continuity equation with the Heisenberg equation for the particle number (see e.g.~Ref.~\cite{Black-Schaffer2008}). In Fig.~3 we plot the disorder averaged clock-wise surface currents along each four sides of an island.
\begin{figure}[b]
\includegraphics[scale = 0.97]{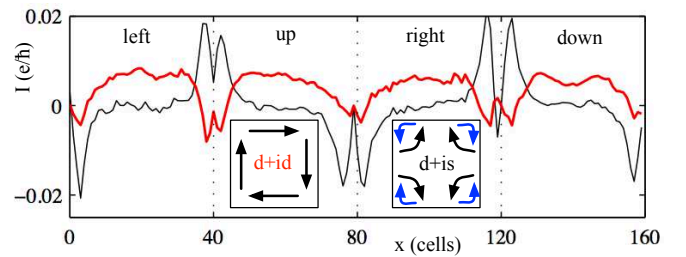}
\caption{\label{fig:current} (Color online) Disorder averaged surface currents in the clockwise direction summed over the 10 first surface layers on each four sides (left, up, right, down) for $L = 40$ and $T_{c,d_2} = 0.5T_c$ (thick red) and $T_{c,s} = 0.5T_c$ (black). Vertical dashed lines mark the corners. Insets show schematically the surface current orientation. Black arrows indicate currents present after disorder averaging, blue arrows indicate outermost surface currents prone to cancelation in disorder averaged results.
}\end{figure}
The $d_1+id_2$ solution has a circulating surface current. While spontaneous, the current is not quantized, in agreement with $d_1+id_2$ superconducting graphene \cite{Black-Schaffer2012}. For $d_1+is$ the surface current instead closes in small separate loops that form a staggered pattern of clockwise and anticlockwise
current flow, with directions displayed in the inset. These localized current vortices will significantly reduce the magnetic field associated with the spontaneous current, consistent with the present experimental situation \cite{Kirtley2006, Saadaoui2011}.

In summary we have shown that the time-reversal symmetry breaking $d_1+is$ state in a cuprate island has a finite energy gap, above which the subsequent energy level spacings are dense. This is very distinct from the $d_1+id_2$ state where the low-energy spectrum is determined by finite size quantization of the two chiral surface states circling the island. In an applied magnetic field, the energy gap increases sub-linearly in the $d_1+is$ state, whereas there are no significant change in energy levels for the $d_1+id_2$ state. Comparing to recent experimental results \cite{Gustafsson2012} our results indicate that a $d_1+is$ state might be present in small YBCO islands.

%
%
\begin{acknowledgments}
We are grateful to A.~V.~Balatsky for discussions. This work was supported by the Swedish research council (VR) and by G\"oran Gustafsson Foundation (ABS).
\end{acknowledgments}


\end{document}